\documentclass[aps,prl,twocolumn,twoside,superscriptaddress,nofootinbib,preprintnumbers,floatfix]{revtex4-2}

\usepackage[bookmarks=false]{hyperref}
\usepackage{graphicx}
\usepackage{amsmath}
\usepackage{mathrsfs}
\usepackage{xspace}
\usepackage{xcolor}
\usepackage[normalem]{ulem}
\usepackage[abs]{overpic}
\usepackage{comment}
\usepackage{dsfont}

\newcommand{\eq}[1]{Eq.~\eqref{eq:#1}}
\newcommand{\eqs}[2]{Eqs.~\eqref{eq:#1} and \eqref{eq:#2}}
\newcommand{\fig}[1]{Fig.~\ref{fig:#1}}
\newcommand{\figs}[2]{Figs.~\ref{fig:#1} and \ref{fig:#2}}

\newcommand{\df}{\mathrm{d}}
\newcommand{\img}{\mathrm{i}}

\newcommand{\tr}{\textrm{tr}}

\newcommand{\de}{\delta}

\newcommand{\si}{\sigma}

\newcommand{\cH}{\mathcal{H}}
\newcommand{\cJ}{\mathscr{J}}

\definecolor{hardcol}{RGB}{239, 187, 47}
\definecolor{beamcol}{RGB}{128, 0, 128}
\definecolor{jetcol}{RGB}{0, 0, 122}
\definecolor{softcol}{RGB}{255, 26, 255}

\newcommand{\nn}{\nonumber}

\allowdisplaybreaks[2]

\DeclareMathOperator{\sech}{sech}
\DeclareMathOperator*{\SumInt}{%
\mathchoice%
  {\ooalign{$\displaystyle\sum$\cr\hidewidth$\displaystyle\int$\hidewidth\cr}}
  {\ooalign{\raisebox{.14\height}{\scalebox{.7}{$\textstyle\sum$}}\cr\hidewidth$\textstyle\int$\hidewidth\cr}}
  {\ooalign{\raisebox{.2\height}{\scalebox{.6}{$\scriptstyle\sum$}}\cr$\scriptstyle\int$\cr}}
  {\ooalign{\raisebox{.2\height}{\scalebox{.6}{$\scriptstyle\sum$}}\cr$\scriptstyle\int$\cr}}
}
\begin{document}

\preprint{UWThPh 2025-3}

\title{$q_T$-slicing with multiple jets}

\author{Rong-Jun Fu}
\email{rjfu23@m.fudan.edu.cn}
\affiliation{Department of Physics and Center for Field Theory and Particle Physics, Fudan University, Shanghai, 200433, China}

\author{Rudi Rahn}
\email{rudi.rahn@univie.ac.at}
\affiliation{University of Vienna, Faculty of Physics, Boltzmanngasse 5, A-1090 Wien, Austria}

\author{Ding Yu Shao}
\email{dingyu.shao@cern.ch}
\affiliation{Department of Physics and Center for Field Theory and Particle Physics, Fudan University, Shanghai, 200433, China}
\affiliation{Key Laboratory of Nuclear Physics and Ion-beam Application (MOE), Fudan University, Shanghai, 200433, China}
\affiliation{Shanghai Research Center for Theoretical Nuclear Physics, NSFC and Fudan University, Shanghai 200438, China}

\author{Wouter J.~Waalewijn}
\email{w.j.waalewijn@uva.nl}
\affiliation{Nikhef, Theory Group,
	Science Park 105, 1098 XG, Amsterdam, The Netherlands}
\affiliation{Institute for Theoretical Physics Amsterdam and Delta Institute for Theoretical Physics, University of Amsterdam, Science Park 904, 1098 XH Amsterdam, The Netherlands}

\author{Bin Wu}
\email{b.wu@cern.ch}
\affiliation{Instituto Galego de F\'isica de Altas Enerx\'ias IGFAE, Universidade de Santiago de Compostela, E-15782 Galicia-Spain}

\begin{abstract}
Modern collider phenomenology requires unprecedented precision for the theoretical predictions, for which slicing techniques provide an essential tool at next-to-next-to-leading order (NNLO) in the strong coupling. The most popular slicing variable is based on the transverse momentum $q_T$ of a color-singlet final state, but its generalization to final states with jets is known to be very difficult. Here we propose two generalizations of $q_T$ that can be used for jet processes, providing proof of concept with an NLO slicing for $pp \to 2$ jets. We present factorization formulae that enable our approach to NNLO, calculate the NNLO collinear-soft function and demonstrate slicing at this order for $e^+e^- \to 2$ jets. One of these generalizations of $q_T$ only applies to planar Born processes, such as $pp \to 2$ jets, but offers a dramatic simplification of the soft function. We also discuss how our approach can directly be extended to obtain predictions for the fragmentation of hadrons. This presents a promising path for high-precision QCD calculations with multi-jet final states.
\end{abstract}

\maketitle

{\it Introduction. --}
The excellent performance of the LHC experiments has increased the need for precise theoretical predictions. A crucial challenge towards precision is the perturbative corrections due to Quantum Chromodynamics (QCD), particularly for final states involving jets. Here much progress has been made in recent years extending to $pp \to 3$ jets at second order in perturbation theory~\cite{Czakon:2021mjy,Abreu:2021oya,Alvarez:2023fhi} (which was already used in experimental measurements~\cite{ATLAS:2024png}) and even third order for some processes involving jets, e.g.~\cite{Chen:2025utl,Fox:2025cuz,Jakubcik:2024poz,Jakubcik:2022zdi,Mondini:2019gid,Currie:2018fgr}. The highest precision is typically achieved by individual calculations, while automated procedures are available primarily for leading order (LO) and next-to-leading order (NLO). Next-to-next-to-leading order (NNLO) differential cross sections by contrast are only available for relatively simple processes through public codes~\cite{Gavin:2010az,Grazzini:2017mhc,Camarda:2019zyx,Catani:2019hip,Campbell:2019dru,Neumann:2022lft}. 

An important bottleneck is the handling of the cancellation of infrared (IR) divergences between real and virtual diagrams, which is guaranteed by the KLN theorem~\cite{Kinoshita:1962ur,Lee:1964is} but challenging to arrange in practice.
 A range of approaches have been developed~\cite{Gehrmann-DeRidder:2005btv,Catani:2007vq,Czakon:2010td,Boughezal:2011jf,Currie:2013vh,Czakon:2014oma,Boughezal:2015dva,Gaunt:2015pea,Cacciari:2015jma,Sborlini:2016gbr,DelDuca:2016ily,Caola:2017dug,Magnea:2018hab,Herzog:2018ily,TorresBobadilla:2020ekr,Bertolotti:2022aih},
which can be roughly subdivided into local subtraction methods and slicing methods. While local subtractions are numerically more stable, slicing methods are easier to extend to new processes (illustrated by e.g.~the quick succession of results using 1-jettiness discussed below). 
A key challenge for slicing is the choice of a suitable resolution variable. In this Letter we introduce two novel transverse-momentum-based slicing variables for multi-jet final states.

For color-singlet production, its transverse momentum  $q_T = |\vec{q}_T|$  serves as an effective slicing variable~\cite{Catani:2007vq}: the virtual contribution occurs at  $q_T = 0$ while real radiation leads to $q_T > 0$.\footnote{This approach also holds for massive colored particles, like the top quark~\cite{Catani:2019iny}.} Slicing capitalizes on this, by splitting the cross section into pieces without (unres.) and including (res.)  additional resolved emissions. As unresolved real emissions are soft or collinear, the former contribution can then be approximated to leading power (LP) in the slicing cutoff:
\begin{align} \label{eq:slicing}
  \frac{\df \si}{\df X} &= \int_0^\de \! \df q_T\, \frac{\df \si_{\rm unres.}}{\df X\, \df q_T}
   + \int_\de^\infty \! \df q_T\, \frac{\df \si_{\rm res.}}{\df X\, \df q_T} \\ &=\int_0^\de \! \df q_T\, \frac{\df \si_{\rm LP}}{\df X\, \df q_T} [1 + \mathcal{O}(\delta^p)]
   + \int_\de^\infty \! \df q_T\, \frac{\df \si_{\rm res.}}{\df X\, \df q_T}\nn
\,.\end{align}
Typically one uses a factorization formula for the cross section for $q_T < \delta$, obtained e.g.~using Soft-Collinear Effective Theory (SCET)~\cite{Bauer:2000yr,Bauer:2001ct,Bauer:2001yt,Bauer:2002nz,Beneke:2002ph}, to approximate the cross section and handle the cancellation of IR divergences. The contribution for $q_T> \delta$ reduces to a simpler calculation, trading an  $\alpha_s$ from a complicated loop calculation for an $\alpha_s$ for a simpler resolved additional parton in the final state.
In \eq{slicing}, $X$ denotes kinematics of the color-singlet final-state (e.g.~its rapidity).
This factorization is not exact, as indicated by the $\mathcal{O}(\delta^p)$ power corrections. Thus $\de \ll 1$ is required, which however leads to large cancellations between the two terms in \eq{slicing},
affecting the numerical stability. 

For processes with jets, such as  $pp \to Z +$jet or $pp \to 2$ jets, $q_T$ \emph{without modification} is unsuitable as slicing variable because radiation emitted inside a jet leaves $q_T=0$. As an alternative, the  $N$-jettiness variable~\cite{Stewart:2010tn} has been proposed~\cite{Boughezal:2015dva, Gaunt:2015pea}, and successfully applied to processes involving a color-singlet plus one jet in the final state~\cite{Boughezal:2015aha,Boughezal:2015ded,Boughezal:2016dtm,Boughezal:2016isb,Campbell:2019gmd,Mondini:2021nck,Kim:2024kaq}. One challenge in extending this to multi-jet final states is the complicated form of the soft function at NNLO~\cite{Bell:2023yso,Agarwal:2024gws} that enters the factorized cross section $\si_{\rm LP}$. For color-singlet processes, $q_T$ performs better than 0-jettiness as slicing variable~\cite{Campbell:2022gdq}, motivating the search for an extension of $q_T$ to processes with jets. In this context, $k_T$-ness was recently introduced~\cite{Buonocore:2022mle,Buonocore:2023rdw}, but implementing it at NNLO will be very challenging in the absence of a factorization formula.

In this Letter we propose two ways of extending  $q_T$ as slicing variable, to processes with jets, using $pp \to 2$ jets as example. The first extension has a very simple factorization, but is restricted to processes that are planar at LO, while the second can be used in general and converges faster, at the price of a more complicated soft function. The key ingredient is the use of a recoil-free jet axis, e.g.~by employing the winner-take-all (WTA) recombination scheme~\cite{Salam:WTAUnpublished, Bertolini:2013iqa}. 
This seemingly subtle change enables slicing because the jet momentum is now deflected by radiation inside the jet, leading to a non-vanishing $q_T$.
The transverse momentum decorrelation $q_T$ we consider here is the vector sum of the momenta of all identified color-singlets and reconstructed jets in the event\footnote{For a single color singlet $q_T$ is that singlet’s transverse momentum, for a dijet system it is the sum of the two reconstructed jet momenta, and for a system of multiple jets and color singlets it is the sum of all jet and color singlet momenta.}. 
Note that we only need to use this different recombination scheme to obtain our slicing variable $q_T$, and that the jets can be defined in the standard way.
As proof of concept, we demonstrate our new slicing at NLO for $pp \to 2$ jets and NNLO for $e^+e^- \to 2$ jets. We present factorization formulae that can readily be used to extend $pp \to$ multiple jets to NNLO, discuss the ingredients that enter in it, and calculate the NNLO collinear-soft function (discussed in the supplemental material~\cite{supplemental}).

\begin{figure}
\includegraphics[width=0.35\textwidth]{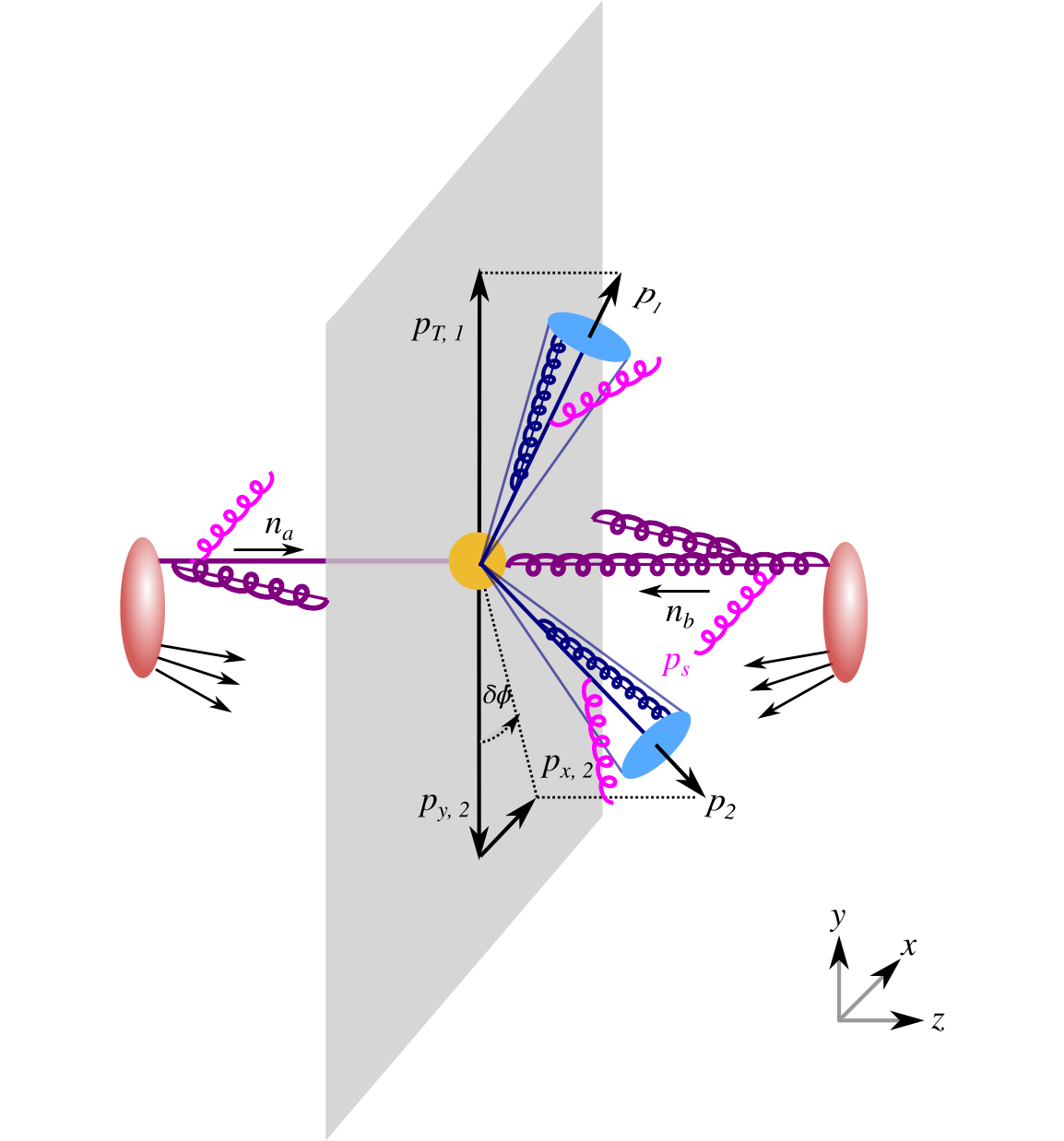}
\caption{The $pp \to 2$ jet process. By using the WTA scheme, the transverse momentum perpendicular to the scattering plane $q_x$ (equal to $p_{x,2}$ in the above coordinates), or equivalently the azimuthal decorrelation $\delta \phi$, is a suitable slicing variable. The ingredients in the corresponding factorization are: the hard scattering (yellow), collinear initial- (purple) and final-state (blue) radiation, and soft radiation (pink).
\label{fig:coordinates}}
\end{figure}

{\it $q_x$ with jets: the planar case. --}
%
For planar Born processes, such as $pp \to H$+jet or $pp \to 2$ jets, we can use the transverse momentum component $q_x$ \emph{perpendicular} to this scattering plane as slicing variable, or equivalently the azimuthal decorrelation $\delta \phi$, see \fig{coordinates}. By using the WTA scheme, the jet axis is also affected by radiation inside the jet leading to a nonzero $q_x$ (one of the components of $\vec q_T$), making it suitable as slicing variable.

The key selling point of $q_x$ as slicing variable is that we have a factorization formula with particularly simple ingredients. Denoting the transverse momenta and rapidities of the jets with $p_{T,1}, \eta_{1}, \eta_{2}$, and building on ~\cite{Chien:2020hzh, Chien:2022wiq}, the cross section for small $q_x$ factorizes as
\begin{align} \label{eq:fact_qx}
   &\frac{\df \si_{\rm LP}}{\df p_{T,1}\, \df \eta_{1}\, \df \eta_{2}\, \df q_x}
   \\
   &= \int\! \frac{\df b_x}{2\pi}\, e^{\img q_x b_x} \sum_{i,j,k,\ell} \textcolor{beamcol}{B_i(x_a,b_x)\, B_j(x_b,b_x)}    \textcolor{jetcol}{\cJ_k(b_x) \cJ_\ell(b_x)}
   \nn \\ & \quad \times
   \tr\bigl[ \textcolor{hardcol}{\hat \cH_{ij \to k\ell}(p_{T,{1}}, \eta_{1} - \eta_{2})}\, \textcolor{softcol}{\hat S_{ijk\ell}(b_x, \eta_{1}, \eta_{2})}\bigr]
\,.\nn\end{align}
This involves the \emph{standard} TMD beam functions $B_{i,j}$, whose matching onto parton distribution functions is well-known at NNLO~\cite{Catani:2011kr,Catani:2012qa,Gehrmann:2012ze,Gehrmann:2014yya,Luebbert:2016itl,Echevarria:2016scs,Luo:2019hmp,Luo:2019bmw}, a soft function that can directly be obtained~\cite{Gao:2019ojf} from the \emph{standard} TMD soft function at NNLO~\cite{Echevarria:2015byo,Lubbert:2016rku}\footnote{The additional tripole contribution to the NNLO soft function~\cite{Gao:2023ivm} drops out when combined with the LO hard function}, and TMD jet functions that are already partially known at NNLO~\cite{Gutierrez-Reyes:2019vbx,Fang:2024auf} and will soon be available fully~\cite{Bell:2021dpb,Brune:2022cgr}\footnote{In principle, the jet function depends on the jet radius $R$, but this drops out for $\delta \phi/R \ll 1$. The only missing ingredient is currently the two-loop constant for the gluon jet function.}. In particular, the analytic dependence of the soft function for this observable on the jet kinematics (see eq.~(8) of \cite{Gao:2019ojf}), which enters through a boost in the scattering plane to make two of the beams/jets back-to-back, should be contrasted with the much more complicated kinematic dependence of the 2-jettiness soft function that requires numerical methods~\cite{Bell:2023yso,Agarwal:2024gws}. The reason why the soft function is so simple in this case, is that the recoil from soft radiation is independent of the region of phase space it is emitted into. 
Note that one additional advantage of recoil-free schemes is that they typically remove~\cite{Neill:2016vbi} so-called non-global logarithms~\cite{Dasgupta:2001sh} from the soft function, which here simplifies the structure of the factorization theorem~\cite{Gutierrez-Reyes:2018qez,Chien:2022wiq}.

The hard function $\hat \cH$ in \eq{fact_qx} describes the hard partonic scattering process $i j \to k \ell$, which for $pp\to2$ jets has been obtained at NNLO~\cite{Broggio:2014hoa} from color-decomposed helicity amplitudes~\cite{Bern:2002tk,Bern:2003ck,Glover:2003cm,Glover:2004si,DeFreitas:2004kmi}.  $\hat \cH$ and $\hat S$ are matrices in color space (as indicated by the hat) and the trace is over color. The momentum fractions $x_{a,b}$ in the beam functions can be expressed in terms of the jet transverse momenta and rapidities.
Beyond NLO, linearly-polarized contributions to the beam and jet functions must be included~\cite{Catani:2010pd,Chien:2020hzh}.

\begin{figure}[t]
    \centering
        \centering
        \includegraphics[width=\linewidth]{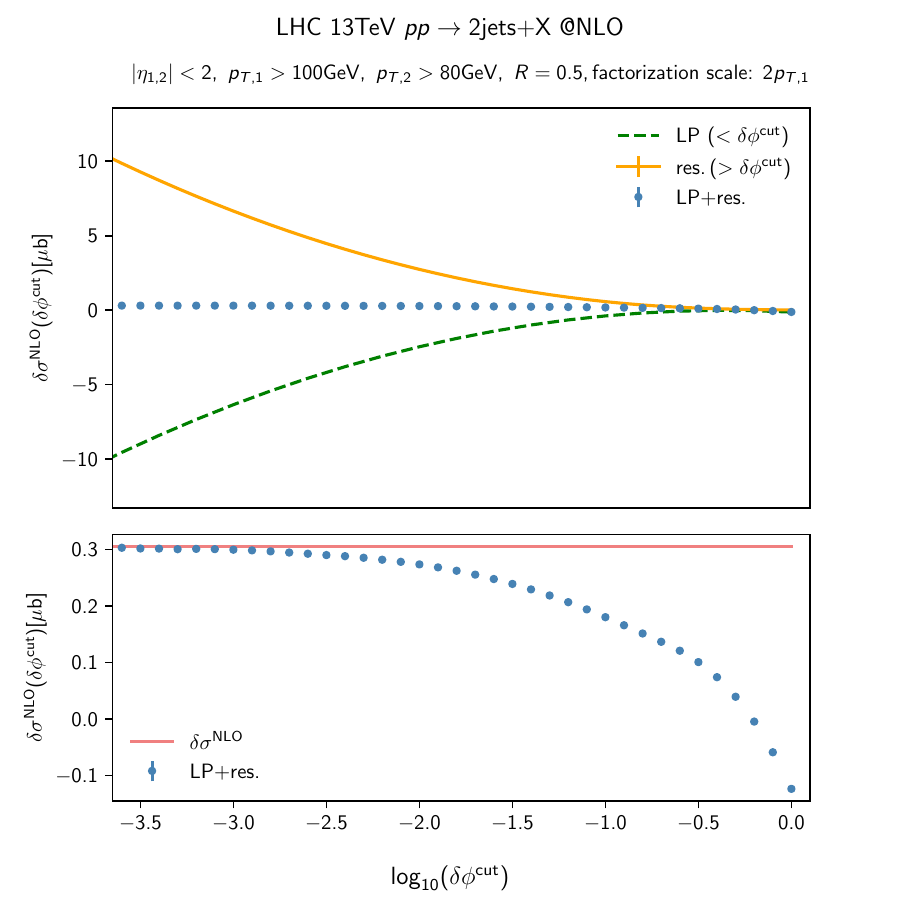}

    \caption{In the lower panel the NLO correction $\delta\sigma^{\rm NLO}$ (blue dots) obtained using the slicing is plotted as a function of the cut on the azimuthal angle $\delta\phi^{\rm cut}$, showing that this converges  for small $\delta\phi^{\rm cut}$ to the correct result (red line) obtained from \texttt{NLOJET++}~\cite{Nagy:2001xb}. In the upper panel the individual terms (green dashed and yellow solid lines) in \eq{slicing} are shown, of which the blue dots are the sum.
    Jets are defined using the anti-$k_T$ algorithm \cite{Cacciari:2008gp} with radius $R = 0.5$, and subject to the following cuts on their transverse momentum and rapidity $p_{T,1}>100$ GeV, $p_{T,2}>80$ GeV, $|\eta_{1,2}|<2$. The renormalization and factorization scales are set to $\mu_{R,F} = 2p_{T,1}$. 
     Note the different ranges of the vertical axis for the two panels, and that the numerical uncertainties are smaller than the size of the markers.}
    \label{fig:qx}
\end{figure}

As proof of concept, we show the result of using $q_x$ as a slicing variable to obtain the NLO correction to dijet cross section $\delta\sigma^{\rm NLO}$ in \fig{qx}. We reiterate that the WTA scheme is only used in the definition of $q_x$, and that the jets themselves are defined in the standard scheme.
As is clear from the bottom panel of \fig{qx}, for small values of $q_x$ this reproduces the NLO cross section. However, as the top panel indicates, there is also a large cancellation between the two terms. This highlights the main bottleneck of slicing (in general), namely the need for numerically precise results for the $\df \sigma_{\rm res.}$ term in \eq{slicing}. 

\begin{figure}[t]
    \centering
        \centering
        \includegraphics[width=\linewidth]{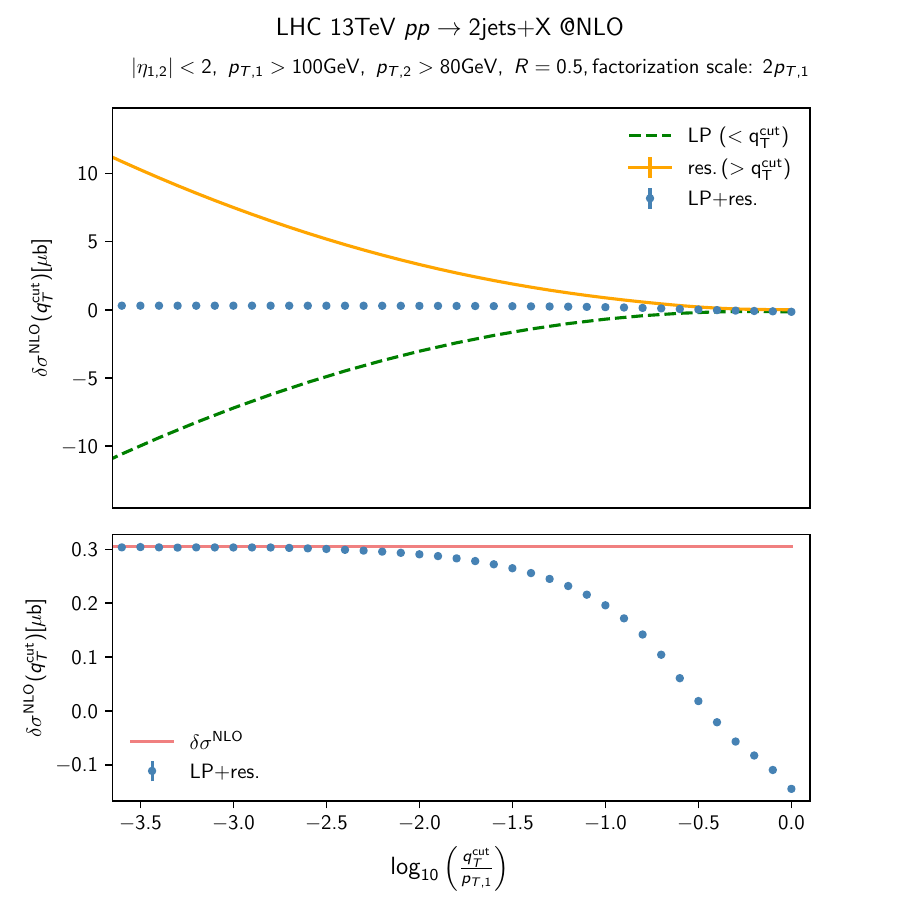}

    \caption{Same as \fig{qx} but using instead a cut on the total transverse momentum $q_T^{\rm cut}$ (with the WTA scheme) for the slicing.
    This converges faster than $\delta \phi^{\rm cut}$ at the expense of a more complicated soft function, and can also be extended to nonplanar Born processes. Note again the different ranges of the vertical axis for the two panels.
    }
    \label{fig:qT}
\end{figure}

{\it $q_T$ with jets: the planar case. --}
For processes that are non-planar at leading order, such as $pp \to Z+2$ jets or $pp \to 3$ jets, the azimuthal decorrelation cannot be used anymore, but the total transverse momentum $q_T = |\vec q_T|$ of the jets (and color-singlet, if present) is still viable as a slicing variable, when using the WTA scheme. This is because additional radiation beyond the LO process again lifts the momentum balance of the final state jet axes, but in general there is no preferred scattering plane. To study the effects of this change, we first analyze the planar case again, now with $q_T$ as slicing variable instead of $q_x$. For dijet production, employing $q_T$ modifies the factorization in \eq{fact_qx} to
\begin{align} \label{eq:fact_qT}
   &\frac{\df \si_{\rm LP}}{\df p_{T,1}\, \df \eta_{1}\, \df \eta_{2}\, \df q_T}
   \\
&= q_T \! \int\!\! \frac{\df^2 \vec b_T}{2\pi}\, J_0(q_T |\vec b_T|) \!\sum_{i,j,k,\ell}\! \textcolor{beamcol}{B_i(x_a,\vec b_T)\, B_j(x_b, \vec b_T)}    \textcolor{jetcol}{\cJ_k(b_x)}
   \nn \\ & \quad \! \times \!
   \textcolor{jetcol}{\cJ_\ell(b_x)}\, \tr\bigl[\textcolor{hardcol}{ \hat \cH_{ij \to k\ell}(p_{T,{1}}, \eta_{1} \!\!-\! \eta_{2})}\, \textcolor{softcol}{\hat S_{ijk\ell}(\vec b_T, \eta_{1}, \eta_{2},R)}\bigr].
\nn\end{align}
Though we have switched from one to two components of the transverse momentum, the only major change compared to \eq{fact_qx} is in the soft function. 

Soft radiation always contributes through momentum conservation, recoiling the collinear radiation. However, if it is emitted into the jet it also contributes through the magnitude of the jet's transverse momentum, which for the WTA scheme is given by the scalar sum. We showed in~\cite{Chien:2022wiq} that to leading power these two contributions cancel each other for the transverse momentum component \emph{parallel} to the jet's transverse momentum, while the \emph{perpendicular} component is unaffected.
Thus outside the jet (where the soft radiation \emph{only} contributes via recoil) soft radiation contributes its full transverse momentum to $\vec q_T$, while inside the jet (where both magnitude and recoil effects are present) only the component perpendicular to the jet axis does. The corresponding soft function knows about the jet,  in particular its radius, making it  more complex than the $R$-independent soft function for $q_x$. A small computational trick allows us to derive analytic instead of numerical results nevertheless: In the limit of small $R$, it refactorizes into simpler global and collinear-soft contributions~\cite{Chien:2015cka}, presented in the supplemental material. The leading term in the small $R$ limit is sufficient to establish the consistency of the factorization in terms of the anomalous dimensions, and in our numerical results we add finite terms up to $\mathcal{O}(R^4)$, achieving sub-percent accuracy on the cross section for $R=0.5$. 

\begin{figure}
\includegraphics[width=0.48\textwidth]{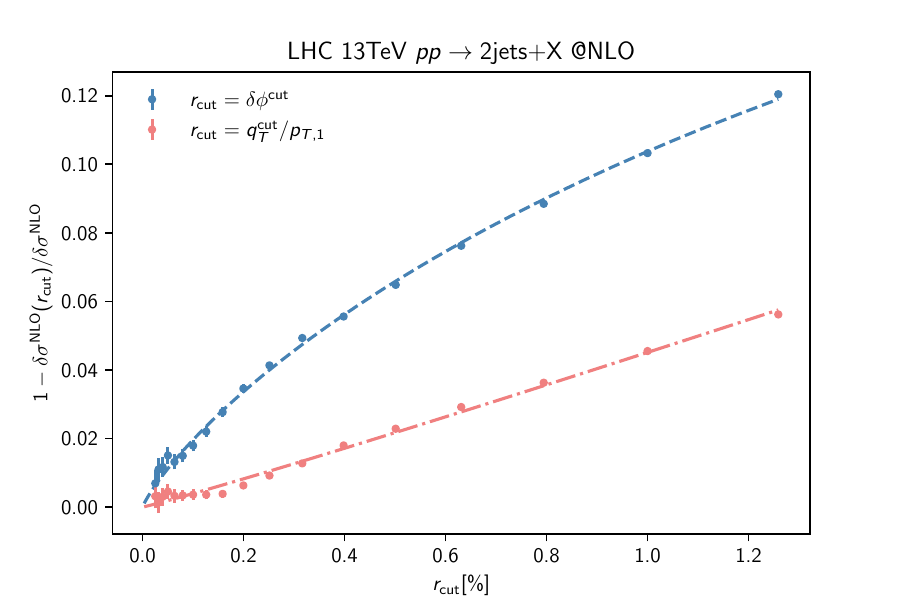}
\caption{A comparison of the precision of the slicing with $r_{\rm cut} = \delta\phi^{\rm cut}$ and $q_T^{\rm cut}/p_{T,1}$ for $pp \to 2$ jets with the same kinematics as in \figs{qx}{qT}. Zooming in on the region of small $r_{\rm cut}$, this clearly shows the faster convergence of $q_T$.
The curves are obtained from fitting to $a\, r_{\rm cut} \ln r_{\rm cut} + b\, r_{\rm cut}$. 
\label{fig:rcut}}
\end{figure}

In \fig{qT} we show that using the total transverse momentum $q_T$ of the two jets works well as a slicing variable. It converges faster than $q_x$ shown in \fig{qx}, but at the price of a more complicated soft function. This faster convergence is particularly clear when comparing the error on the slicing for small values of the slicing variable, shown in \fig{rcut}.

{\it $q_T$ with jets: the nonplanar case. --}
To illustrate that $q_T$ can also be extended to the nonplanar case, we present the factorization formula for $pp \to$ 3 jets: 
\begin{align} \label{eq:fact_3jets}
   &\frac{\df \si_{\rm LP}^{pp \to {\rm 3 jets}}}{\df p_{T,1}\, \df p_{T,2}\, \df \eta_{1}\, \df \eta_{2}\, \df \eta_{3}\, \df \Phi\, \df q_T}
   \\
&= q_T \int\! \frac{\df^2 \vec b_T}{2\pi}\, J_0(q_T |\vec b_T|) \sum_{i,j,k,\ell,m} B_i(x_a,\vec b_T)\, B_j(x_b, \vec b_T)       \nn \\ & \quad  \times 
   \cJ_k(b_{\perp,1})\,\cJ_\ell(b_{\perp,2}) \,\cJ_m(b_{\perp,3})\, 
   \nn \\ & \quad \times   
   \tr\bigl[ \hat \cH_{ij \to k\ell m}(\{p_{T,i}\},\Phi,\{\eta_i\})\, 
   \hat S_{ijk\ell m}(\vec b_T, \{\eta_i\},\Phi,R)\bigr],
\nn\end{align}
which now also depends explicitly on the azimuthal angle $\Phi$ between two jets.
We have verified the consistency of this factorization in terms of anomalous dimensions.
The new challenge is that the jets are no longer in the same transverse plane, so the transverse momentum component perpendicular to the jet depends on the jet at hand. For the jet function this change is minor as only its argument is modified to $b_{\perp,i}$, which is the transverse component perpendicular to the $i$-th jet direction $\hat n_{T,i}$, i.e.~$\vec{b}_{\perp,i} = \vec b_T - \hat n_{T,i} (\hat n_{T,i} \cdot \vec b_T)$. In the soft function the modification is less trivial, particularly for the jet-jet dipoles, as discussed briefly in the supplemental material.

\begin{figure}[htbp]
    \centering
    \includegraphics[width=\linewidth]{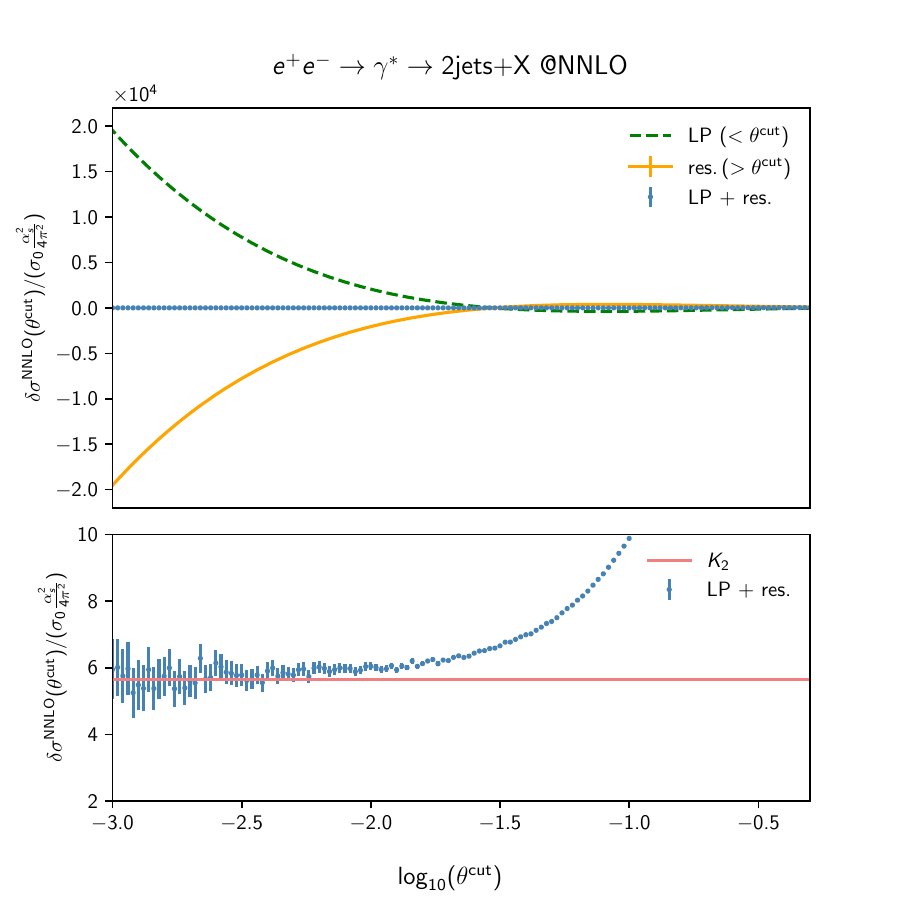}

    \caption{In the lower panel the NNLO correction $\delta\sigma^{\rm NNLO}$ (blue dots with error bars) obtained using the slicing is plotted as a function of $\theta^{\rm cut}$. This converges for small $\theta^{\rm cut}$ to the correct result $K_2$ (red line), given in \eq{K12}. In the upper panel the individual terms (green dashed and yellow solid curves) in \eq{slicing} are shown, of which the blue dots are the sum.
    \label{fig:NNLO_ee}}
\end{figure}

{\it Extension to NNLO --}
While one crucial ingredient for WTA-$q_T$ slicing at hadron colliders is missing, namely the gluon WTA jet function, all ingredients for an implementation at lepton colliders are available. Extending our slicing setup to calculate the inclusive dijet production cross section (for easy comparison to known results), we observe again the expected cancellation between LP and resolved emission contribution, as shown in \fig{NNLO_ee}. We relegate the details of this calculation (and in particular the differences to the hadronic setup) to the supplemental material.

{\it Extension to fragmentation. --}
We can also use the same slicing variables based on transverse momentum to obtain predictions for the fragmentation of hadrons. For planar Born processes, the azimuthal decorrelation can be used again: One simply replaces one of the TMD jet functions in \eq{fact_qx} with a TMD fragmentation function, whose matching onto fragmentation functions is known at NNLO~\cite{Echevarria:2015usa,Echevarria:2016scs,Luo:2019hmp,Luo:2019bmw}.\footnote{Since the hadron only carries a fraction $z_h$ of the momentum of the initial parton, we need to divide the hadronic transverse momentum by $z_h$ to obtain the partonic transverse momentum that enters in the slicing variable.}  If instead of $q_x$ the full transverse momentum $q_T$ is used, the soft function is not the same as in \eqs{fact_qT}{fact_3jets}, since there is no jet algorithm (or dependence on the jet radius) for hadrons.

{\it Conclusions and outlook. --}
In this Letter we have presented two generalizations of the total transverse momentum $q_T$, that are promising slicing variables for LHC processes with multiple jets.
For planar Born processes, like $pp \to 2$ jets, the azimuthal decorrelation $\delta \phi$ can be used to handle the cancellation of IR divergences, while in  general the total transverse momentum $q_T$ can be employed. The key innovation that makes these suitable slicing variables is the use of a recoil-free axis, allowing us to also resolve emissions inside jets.

We demonstrated these slicing variables for $pp \to 2$ jets at NLO, and $e^+e^- \to 2$ jets at NNLO. We
presented the factorization formulae that enable this for general processes with jets at NNLO. Most NNLO ingredients are already available, except for the constant in the gluon jet function, and the soft function for $q_T$. For $\delta \phi$ the NNLO soft function is known and clearly much simpler than for 2-jettiness. Though the power corrections are larger than for $q_T$, we expect that this simplicity should extend to the power corrections. For $q_T$, the NNLO soft function will be of comparable complexity as the $N$-jettiness soft function. However, we expect that the refactorization due to the expansion in $R$ will lead to simplifications, and present the result for the NNLO collinear-soft function in the supplemental material.

The factorization formulae for multi-jet final states can also be used as a baseline for studying factorization violating effects, where there has been significant interest in the transverse momentum of two back-to-back jets, see e.g.~\cite{Collins:2007nk, Rogers:2010dm}.

Overall, this work outlines a promising pathway for achieving high-precision QCD calculations in multi-jet final states, paving the way for further advancements in theoretical and experimental high-energy physics.

{\it Acknowledgements --}
We thank Yang-Ting Chien for collaboration in the early stages of this project and  Hai-Tao Li for helpful discussions.
R.J.F. and D.Y.S. are supported by the National Science Foundations of China under Grant No.~12275052, No.~12147101. D.Y.S. is also supported by the Innovation Program for Quantum Science and Technology under grant No. 2024ZD0300101.
R.R.~is supported by the European Union’s
Horizon Europe research and innovation programme under the Marie
Sk\l{}odowska-Curie project ``SoftSERVE-NGL'' with grant agreement No.
101108359, and acknowledges prior funding through the Royal Society grant
URF\textbackslash{}R1\textbackslash{}201500.
B.W.~is supported by the European Research Council project ERC-2018-ADG-835105 YoctoLHC; by the project CEX2023-001318-M financed by MCIN/AEI/10.13039/-
501100011033; by the Spanish Research State Agency under projects PID2020-119632GBI00 and PID2023-152762NB-I00;  by Xunta de Galicia under the ED431F 2023/10 project; by Xunta de Galicia (Centro singular de investigaci\'on de Galicia accreditation 2019-2022); and by the European Union ERDF.

\bibliography{jet.bib}

\clearpage
\appendix
\onecolumngrid
\setcounter{equation}{0}
\renewcommand{\theequation}{S-\arabic{equation}}

\section*{Supplemental material}
\subsection*{$\vec{q}_T$ soft function}
The soft function in impact parameter space $\vec b_T$ is given by the following correlator 
\begin{align} \label{eq:soft_def}
     \hat S\bigl(\vec{b}_T, \eta_1,\eta_2, R \bigr) &= \SumInt_X \langle 0| \overline{\mathbf T}[
     \sum_n \Theta_n(X) O_s^\dagger(b_n)
     | X \rangle 
    \langle X| \mathbf{T}[O_s(0)]|0\rangle \nn \\
    &= \hat{1} + \hat S^{(1)} \bigl(\vec{b}_T, \eta_1, \eta_2, R\bigr) + \mathcal{O}(\alpha_s^2)
\end{align}
of soft Wilson lines $S_{n_i}$ tracing the beams and jets
\begin{equation}
    O_s(x^\mu) = \bigl[S_{n_2}S_{n_1}S_{n_b}S_{n_a}\bigr](x^\mu)\,.
\end{equation}
The implicit dependence on the jet rapidities $\eta_{1,2}$ is encoded in the light-like vectors $n_i$ along the beam ($i=a,b$) and jet ($i=1,2$) directions, 
$$
n_a^\mu = (1,0,0,1),\quad n_b^\mu = (1,0,0,-1),\quad 
n_1^\mu = (1,0,\operatorname{sech}{\eta_1},\tanh{\eta_1}),\quad n_2^\mu = (1,0,-\operatorname{sech}{\eta_2},\tanh{\eta_2}).
$$
For definiteness, we choose the planar beam-jet system to align with the $yz$-plane. (We will comment on non-planar case below, and note that for each additional jets  an extra Wilson line must be included.)
The indices ``$ijkl$'' of the soft function appearing in the factorization formulae label the color representation $\mathbf{T}_i$ of the corresponding Wilson line. The beam Wilson lines $S_{n_a}$ and $S_{n_b}$ are defined as 
\begin{equation}
    S_{n_i}(x^\mu) = \mathbf{P} \exp{\biggl[\img g \int_{-\infty}^0\mathrm{d} t~ n_i\cdot A_s^c(x^\mu + tn_i^\mu)\, \mathbf{T}^c_i
 \biggr]}\,,
 \label{eq:wilsonline}
\end{equation}
with corresponding expressions for the outgoing Wilson lines for the jets.

In \eq{soft_def}, $\Theta_n(X)$ depends on the various possibilities of soft particles in $X$ in/outside of the jet. At NLO, there is a single emission that is either in or outside the jet, with
\begin{align}
    \Theta_{\text{in}} &= \Theta(R-\Delta R_1) + \Theta(R - \Delta R_2) \nn \\
    &= \Theta\bigl(R - \sqrt{(\phi-\phi_1)^2 + (\eta-\eta_1)^2}\bigr) + \Theta\bigl(R - \sqrt{(\phi-\phi_2)^2 + (\eta-\eta_2)^2}\bigr),
    \label{eq:PSconstrain}
\end{align}
where $R$ represents the jet radius, and $\phi$ and $\eta$ denote the azimuthal angle and pseudo-rapidity of the soft emission momentum, respectively. The azimuthal angles for the two jets are taken to be $\phi_1=0$ and $\phi_2=\pi$. The out-of-jet constraint is complementary to the in-jet constraint: $\Theta_{\text{out}} = 1 - \Theta_{\text{in}}$. Thus at this order, $\sum_n \Theta_n(X) O_s^\dagger(b_n) =  \Theta_{\rm in}(X) O_s^\dagger(b_{\rm in}) +  \Theta_{\rm out}(X) O_s^\dagger(b_{\rm out})$, with
\begin{equation} \label{eq:b_i}
b_{\rm in} = (0,b_x,0,0)\,, \qquad b_{\rm out} = (0,b_x,b_y,0)=(0,\vec{b}_T,0)
\,.\end{equation}
The peculiar form of $b_{\rm in}$ is due to the fact that only the  component of the transverse momentum perpendicular to the jets, here $q_x$, is measured inside the jet. Below we will also employ a polar representation for $\vec{b}_T$ such that $b_x=b_T \sin\phi_b$ and $b_y=b_T \cos\phi_b$. 

\medskip

The explicit form of the next-to-leading order soft function is given by the sum over color dipoles:
\begin{align}\label{eq:Sdip}
    \hat S^{(1)}\bigl(\vec{b}_T, \eta_1, \eta_2, R,\mu,\nu\bigr) 
    &= \sum_{i<j}\bigl(-\mathbf{T}_i\cdot\mathbf{T}_j\bigr) \frac{\alpha_s (\mu)}{4\pi}~ S^{(1)}_{ij}(\vec{b}_T, \eta_1, \eta_2, R,\mu,\nu\bigr)  \notag \\
    &=\sum_{i<j}\bigl(-\mathbf{T}_i\cdot\mathbf{T}_j\bigr) \frac{\alpha_s(\mu)\mu^{2\epsilon}\pi^{\epsilon}e^{\gamma_E\epsilon}}{\pi^2}
    \int \mathrm{d}^d k\, \delta(k^2)\theta(k^0) \left(\frac{\nu}{2k^0}\right)^\eta\frac{n_i\cdot n_j}{n_i\cdot k ~ n_j\cdot k} \notag \\
    &\quad \times \bigl[\Theta_{\text{out}}e^{\img \vec{k}_T\cdot\vec{b}_T}  + \Theta_{\text{in}} e^{\img k_x b_x} \bigr]. 
\end{align}
Using the relation $\Theta_{\text{out}} = 1 - \Theta_{\text{in}}$, the soft function can now be split into three components: a global soft function, as well as a 2-dimensional and 1-dimensional collinear-soft function, 
\begin{align} \label{eq:NLOcoft_def}
    \hat S^{(1)}\bigl(\vec{b}_T, \eta_1, \eta_2, R,\mu,\nu \bigr)
    &=\sum_{i<j}\bigl(-\mathbf{T}_i\cdot\mathbf{T}_j\bigr) \frac{\alpha_s(\mu)\mu^{2\epsilon}\pi^{\epsilon}e^{\gamma_E\epsilon}}{\pi^2}
    \int \mathrm{d}^d k\, \delta(k^2)\theta(k^0) \frac{n_i\cdot n_j}{n_i\cdot k ~ n_j\cdot k} \\
    &\quad \times \biggl[\Bigl(\frac{\nu}{2k^0}\Bigr)^\eta e^{i\vec{k}_T\cdot\vec{b}_T} - \Theta_{\text{in}} e^{i\vec{k}_T\cdot\vec{b}_T} + \Bigl(\frac{\nu}{2k^0}\Bigr)^\eta \Theta_{\text{in}} e^{ik_x b_x} \biggr] \nn \\
    &=  \sum_{i<j}\bigl(-\mathbf{T}_i\cdot\mathbf{T}_j\bigr) \frac{\alpha_s (\mu)}{4\pi}~ \Bigl[S^{\text{global}}_{ij}(\vec{b}_T, \eta_1, \eta_2, \mu, \nu) + S^{cs,2d}_{ij}(\vec{b}_T,R,\mu) + S^{cs,1d}_{ij}(b_x, \eta_1,\eta_2,R, \mu,\nu)\Bigr]\,.
\nn\end{align}
Intuitively, the global soft function measures the out-of-jet measurement everywhere, the 2d collinear-soft function subtracts the incorrect $\vec{q}_T$ measurement inside the jets, and the 1d collinear-soft function adds the correct $q_x$ measurement.
Note that depending on the dipole under consideration the global soft function may not always require the rapidity regulator.
The dependence of the two collinear-soft functions on the jet radius can be obtained analytically, by expanding around $R=0$. We need to be mindful of the fact that we assumed for the jet function that $R\gg q_T/Q$, which means we typically need to include the subleading dependence in $R$ in either the soft function (or the jet function instead). This allows us to derive analytic formulae, and is a mere computational tool: Our aim is not to investigate any $R$-dependency in logarithms, for us it is just a parameter.

In this Letter we are only interested in the magnitude of the transverse momentum imbalance $q_T$ between the two jets, so we average the soft function's components over the azimuthal angle $\phi_b$ of $\vec{b}_T$, which we indicate with a bar.

The global soft function (for the 2-component vector $\vec{b}_T$) is known~\cite{Kang:2021ffh,delCastillo:2020omr}, we thus merely have to compute the two collinear-soft functions explicitly. After Fourier transforming from $\vec{b}_T$ space to $\vec{q}_T$ space and integrating $q_T$ up to $q_T^\text{cut}$, we obtain: 
\begin{align}
    \hat{\bar{S}}^{(1)}(q_T^{\text{cut}}, \eta_1, \eta_2, R, \mu,\nu) &= 
    \int \mathrm{d}^2\vec{q}_T\,  \Theta(q_T^{\rm cut} - q_T) \int \frac{\mathrm{d}^2 \vec{b}_T}{(2\pi)^2}\,  e^{-i\vec{b}_T\cdot \vec{q}_T} \hat{S}^{(1)}\bigl(\vec{b}_T, \eta_1, \eta_2, R, \mu,\nu \bigr) 
    \nn \\
    &= q_T^{\rm cut} 
   \int_0^\infty \mathrm{d}b_T J_1(b_T q_T^{\rm cut})\, \hat{\bar{S}}^{(1)}(b_T, \eta_1, \eta_2, R, \mu,\nu),
\end{align}
where $J_1$ is a Bessel function.

Renormalizing to subtract the regulator poles yields the following finite result, expanded around the small-$R$ limit:
\begin{align}
    \hat{\bar{S}}^{(1)}_{\text{finite}}(q_T^{\text{cut}}, \eta_1, \eta_2, R,\mu,\nu) &= \frac{\alpha_s(\mu)}{4\pi}\,\biggl\{
    -4L_\mu^2\sum_i \mathbf{T}_i^2  + L_\mu\biggl[4\ln\frac{\mu^2}{\nu^2} \sum_i\mathbf{T}_i^2
    +8\sum_{i<j}\mathbf{T}_i\cdot\mathbf{T}_j \ln\frac{n_i\cdot n_j}{2} \nn \\
    &\quad -8\ln2~(\mathbf{T}_a + \mathbf{T}_b)\cdot(\mathbf{T}_1 + \mathbf{T}_2) - 16\ln2~\mathbf{T}_1 \cdot \mathbf{T}_2\biggr]  - \frac{\pi^2}{6}\sum_i \mathbf{T}_i^2 \nn \\
    &\quad + \left[(\mathbf{T}_a + \mathbf{T}_b)\cdot(\mathbf{T}_1 + \mathbf{T}_2) + 2~\mathbf{T}_1 \cdot \mathbf{T}_2\right]
    \Bigl(4\ln2\ln\frac{\mu^2}{\nu^2} + \frac{\pi^2}{3} + 4\ln^2\frac{R}{2}
    \Bigr) \nn \\
    &\quad +\sum_{j\in\text{jets}} (\mathbf{T}_a + \mathbf{T}_b)\cdot\mathbf{T}_j~8\ln2\ln{(2\cosh{\eta_j})} \nn \\
    &\quad +\mathbf{T}_1\cdot\mathbf{T}_2 \left[ 8\ln2\ln(4\cosh{\eta_1}\cosh{\eta_2}) - 2\ln^2(2+2\cosh(\eta_1-\eta_2)) + 2(\eta_1-\eta_2)^2
    \right] \nn \\
    &\quad - \sum_{i<j}\mathbf{T}_i\cdot \mathbf{T}_j \, S^{\text{corr}}_{ij}(\eta_1,\eta_2,R) \biggr\}
    \label{eq:soft_finiteNLO}
\end{align}
where $L_\mu = \ln(\mu/q_T^\text{cut})$ and the sum on $i,j$ run over the ordered set $\{a,b,1,2\}$, with $a,b$ the beams, and $1,2$ the jets. The subleading $R$-correction terms for the various dipoles are given by:
\begin{equation}
    S^{\text{corr}}_{ab}(\eta_1,\eta_2,R) = 4R^2\left(1 - 2\ln{\frac{R}{2}}\right) + \frac{R^4}{3} +\mathcal{O}(R^6)\,,
\end{equation}
\begin{align}
    S^{\text{corr}}_{a1 \text{ or } b2}(\eta_1,\eta_2,R) = &-R^2 \ln{\frac{R}{2}} \left[1 + \frac{4e^{2\eta_2}}{(e^{\eta_1}+e^{\eta_2})^2}\right]
    + R^2 \left[ \frac{7}{6} + \frac{2e^{2\eta_2}}{(e^{\eta_1}+e^{\eta_2})^2} \right] \nn \\
    &+R^4\left[\frac{49}{1440} + \frac{e^{2\eta_2}  \left( 10e^{2\eta_1} + e^{2\eta_2} - 4e^{\eta_1 + \eta_2} \right)}{6  (e^{\eta_1} + e^{\eta_2})^{4}} 
    - \ln\frac{R}{2}  \left( \frac{1}{72} + \frac{2e^{2(\eta_1 + \eta_2)}}{(e^{\eta_1} + e^{\eta_2})^{4}} \right) \right] 
    +\mathcal{O}(R^6)\,,
\end{align}
\begin{align}
    S^{\text{corr}}_{a2 \text{ or } b1}(\eta_1,\eta_2,R) = &-R^2 \ln{\frac{R}{2}} \left[1 + \frac{4e^{2\eta_1}}{(e^{\eta_1}+e^{\eta_2})^2}\right]
    + R^2 \left[ \frac{7}{6} + \frac{2e^{2\eta_1}}{(e^{\eta_1}+e^{\eta_2})^2} \right] \nn \\
    &+R^4\left[ \frac{49}{1440} + \frac{e^{2\eta_1}  \left( 10e^{2\eta_2} + e^{2\eta_1} - 4e^{\eta_1 + \eta_2} \right)}{6  (e^{\eta_1} + e^{\eta_2})^{4}} 
    - \ln\frac{R}{2}  \left( \frac{1}{72} + \frac{2e^{2(\eta_1 + \eta_2)}}{(e^{\eta_1} + e^{\eta_2})^{4}} \right) \right]
    +\mathcal{O}(R^6)\,,
\end{align}
\begin{align}
    S^{\text{corr}}_{12}(\eta_1,\eta_2,R) =& -2 R^2 \ln{\frac{R}{2}} \tanh^2{\left( \frac{\eta_1-\eta_2}{2} \right)} 
    + R^2 \left[ \frac{7}{3} - 
    \frac{6}{1+\cosh{(\eta_1 - \eta_2)}}\right] \\
    &+R^4 \left[ \frac{49}{720} - \frac{e^{\eta_1 + \eta_2}  \left( 3e^{2\eta_1} + 3e^{2\eta_2} - 8e^{\eta_1 + \eta_2} \right)}{2  (e^{\eta_1} + e^{\eta_2})^{4}} 
     - \ln\left(\frac{R}{2}\right)\frac{\left( e^{2\eta_1} + e^{2\eta_2} - 10e^{\eta_1 + \eta_2} \right)^{2}}{36  (e^{\eta_1} + e^{\eta_2})^{4}} \right] + \mathcal{O}(R^6)\,.
     \nn
\end{align}

This soft function can be extended to the case in which the beams and jets do not form a common plane, or to cases with additional jets. In these situations some subtleties and changes arise, though the calculation of the various ingredients is only marginally modified at NLO:
\begin{itemize}
    \item The leading beam-beam dipole contribution is unaffected, as its global soft contribution is insensitive to the jets, and no collinear-soft contribution is present. 
    \item The beam-jet dipole global soft contributions are largely unaffected, as the impact of the azimuthal angle of the jet amounts to a simple rotation of the plane formed by beam and jet. This can be accounted for by shifting the azimuthal variable $\phi_b$ for the vector $\vec{b}_T$ accordingly. The same applies to the collinear-soft contribution.
    \item For a jet-jet dipole the collinear-soft contribution is the same as for a beam-jet dipole, just now for both jets, and can thus be easily adapted. The global soft contribution on the other hand picks up explicit dependence on the azimuthal separation of the two jets through the dipole kinematics, in addition to any potential global rotation accounted for by a shift in $\phi_b$. This azimuthal separation dependence (which we assumed in this Letter to be equal to $\pi$) marginally modifies the appearing expressions, but does not add any difficulties.
    \item The subleading $R$-corrections in general exhibit a subtlety: They depend on the azimuthal orientation $\phi_b$ of the impact parameter $\vec{b}_T$, over which we average here, after which we expand in $R$. (Without expansion we were unable to find analytic expressions for the appearing terms.) This $R$-expansion however does not commute with the averaging, and more specifically does not commute with evaluations for certain special values of $\phi_b$ (with the $R$-expansion converging badly in their vicinity). If $\phi_b$ is not averaged the expansion should thus not be performed, which means the $R$-corrections must be evaluated numerically (the appearing integrals are either 1- or 2-dimensional).
    \item The $R$-corrections generally distinguish between active jets (those involved in the dipole, and thus already partially included via a collinear-soft function), and passive jets (those in the bulk). The beam-beam case only exhibits passive corrections, while for the jet-jet case the passive corrections can only be relevant if three or more jets are present. In general the passive jet corrections are more complicated than the active ones, as they encode more information about correlations between jets.
\end{itemize}
We have worked out the changes arising in non-planar and multi-jet situations, and while they are beyond the scope of this Letter, they can be made available upon request.

\subsection*{NNLO slicing in dijet production at lepton colliders}

To demonstrate the effectiveness of our slicing variable at NNLO accuracy, we apply our $q_T$-slicing framework to the NNLO calculation of dijet production process in $e^+e^-$ collisions. In this case,  $q_T$ is the momentum component of the subleading jet transverse to the leading jet (for which the factorization is equivalent to that of $q_x$ in hadron collisions).  

Since jets are clustered using the anti-$k_T$ algorithm in hadron colliders, we adopt its generalization for electron-positron colliders, where the distances are defined as follows:
\begin{align} \label{eq:ee_antikt}
    d_{ij} = \min(E_i^{-2}, E_j^{-2}) \frac{1-\cos{\theta_{ij}}}{1-\cos{R}}, \, \qquad d_{iB} = E_i^{-2}.
\end{align}
For our calculation we choose $R=0.5$, and the transverse momentum measurement is again based on a WTA recombination scheme to determine $q_T=|\vec{q}_T|$.
The constant term of the WTA quark jet function at NNLO is taken from the calculation of~\cite{JetFunctionsSiegen}, with the exception of the $C_FC_A$ color structure, for which the uncertainties from numerical integration are sizable. For this color structure we rely on the fit~\cite{Gutierrez-Reyes:2019vbx} using \textsc{Event2}~\cite{Catani:1996vz}. A fit of the total constant (i.e.~summing over all color structures) was also presented in Ref.~\cite{Fang:2024auf}, but with larger uncertainties.

Following Ref.~\cite{Gutierrez-Reyes:2019vbx} we define our slicing variable $\theta$ as
\begin{align} \label{eq:ee_slicing}
\theta = \arctan\!\left(\frac{2q_T}{Q}\right) \approx \frac{2q_T}{Q},
\end{align}
which is equal to $\pi$ minus the angle between the leading and subleading jets.

The perturbative expansion for the inclusive total cross section $\sigma$ of the $e^+e^-\rightarrow \gamma^* \rightarrow$ dijet process is given by:
\begin{align}
    \sigma = \sigma_0 \biggl[ 1 + \frac{\alpha_s(Q)}{2\pi}K_1 + \left(\frac{\alpha_s(Q)}{2\pi}\right)^2 K_2 + \mathcal{O}(\alpha_s^3)\biggr], 
\end{align}
with one- and two-loop coefficients~\cite{Dine:1979qh, Celmaster:1979xr}
\begin{align} \label{eq:K12}
    K_1 = 2,\quad K_2 = \frac{365}{6}-44\zeta_3+n_f\left(\frac{8}{3}\zeta_3-\frac{11}{3}\right).
\end{align}
Here, $\sigma_0$ denotes the LO cross section, $Q$ denotes the center-of-mass energy of the collision, and the number of (light) quark flavors is chosen as $n_f = 5$.

\subsection*{NNLO collinear-soft function}

To highlight another advantage of our slicing scheme, namely its more readily available analytic control over relevant process parameters, we present the collinear-soft contribution at NNLO, for the full $q_T$, non-planar case. For a complete NNLO treatment of this slicing scheme this is on its own not sufficient, it must be combined with the global soft function at NNLO (which is independent of the jet recombination scheme and radius, and thus relevant to other processes, as well), the NNLO WTA jet functions, and subleading jet radius corrections in either the soft or jet sectors.

The explicit form of the bare collinear-soft function for the jet labeled by the final state parton $j$, taken to have azimuthal orientation $\phi=0$, in the small-$R$ limit and in two-dimensional Fourier space $\vec{b}_T$, is given by:
\begin{align} \label{eq:coft_noAvg}
    S_j^{cs}(b_\perp,\,b_-) &= 1 + \frac{Z_\alpha\alpha_s}{4\pi} S_j^{\rm in}(b_\perp) + \frac{Z_\alpha\alpha_s}{4\pi} S_j^{\rm out}(b_-) 
    + \left( \frac{Z_\alpha\alpha_s}{4\pi} \right)^{\!2} S_j^{cs,(2)}(b_\perp,\,b_-)  + \mathcal{O}(\alpha_s^3),
\end{align}
with $\alpha_s=\alpha_s(\mu)$, $b_\perp = b_x = b_T\sin{\phi_b}$ and
$b_- = b\cdot n_j = -b_T\cos{\phi_b}\sech{\eta_j}$.
Note that $n_j$ is the full jet direction, unlike the normalized projection onto the transverse plane $\hat n_{T,j}$ in the main text.  We use $\overline{\rm MS}$ renormalization scheme, 
\begin{align*}
    g_{s,0}^2 = 4\pi\left(\frac{\mu^2e^{\gamma_E}}{4\pi}\right)^{\!\epsilon} \alpha_s Z_{\alpha}, \quad
    Z_{\alpha} = 1 - \frac{\alpha_s}{4\pi} \frac{\beta_0}{\epsilon} + \mathcal{O}\left( \alpha_s^2 \right).
\end{align*}
To simplify the variable dependence in the small $R$ limit, we have suppressed in our notation non-critical variable dependence of the functions appearing here.

To prepare for the upcoming Non-Abelian Exponentiation (NAE) and the subsequent $\phi_b$-averaging procedure, we first present the $\phi_b$-dependent NLO quark collinear-soft function, retaining the positive $\epsilon$ and $\eta$ coefficients, 

\begin{align} \label{eq:NLO_coft}
    S_j^{\rm in}(b_\perp) = C_F \left( \frac{\mu |b_\perp|}{b_0}\right)^{\!2\epsilon} \left( \frac{\nu |b_\perp|R_j}{b_0}\right)^{\!\eta}\,h_F^{\rm in}, \qquad 
    S_j^{\rm out}(b_-) = C_F \left( \frac{i\, b_- \mu}{b_0 R_j}\right)^{\!2\epsilon} \,h_F^{\rm out},
\end{align}
where $b_0=2e^{-\gamma_E},\, R_j = R/(2\cosh{\eta_j})$ and 
\begin{align}
    h_F^{\rm in} &= \frac{2}{\epsilon^2} - \frac{4}{\eta\epsilon} - \frac{\pi^2}{6} 
    +\left(-\frac{\eta}{\epsilon^3} + \frac{\pi^2\eta}{12\epsilon} - \frac{\zeta_3}{3}\eta- \frac{\pi^2\epsilon}{3\eta} - \frac{4\zeta_3}{3}\epsilon - \frac{17\pi^4}{1440}\eta\epsilon - \frac{4\zeta_3\epsilon^2}{3\eta} - \frac{3\pi^4}{80}\epsilon^2 - \frac{\pi^4\epsilon^3}{40\eta} \right) , \\
    h_F^{\rm out} &= -\frac{2}{\epsilon^2} - \frac{\pi^2}{2} +\left(- \frac{14\zeta_3}{3}\epsilon - \frac{7\pi^4}{48} \epsilon^2\right).
\end{align}
This function is the non-averaged version of the NLO collinear-soft function at the beginning of the supplemental material. 

Since the difference between final state quarks and gluons is completely encoded in the color representation of the corresponding Wilson lines,
the NLO gluon collinear-soft function can be obtained by replacing the color factor $C_F$ with $C_A$ in Eq.~\eqref{eq:NLO_coft}. To obtain the NNLO gluon collinear-soft function from Eq.~\eqref{eq:bare_coft_phibavg} below, we will only need to change the color factors $C_F^2$ and $C_F C_A$ to $C_A^2$ and $C_F n_f T_F$ to $C_A n_f T_F$. 

For slicing at NNLO we only require the $\phi_b$-averaged collinear-soft function, which we organize as follows:
\begin{align}
  \bar{S}_j^{cs,(2)}(b_T) = \int_{0}^{2\pi} \frac{\df \phi_b}{2\pi} S_j^{cs,(2)}(b_\perp,\,b_-)
\end{align}
\begin{align} \label{eq:bare_coft_phibavg}
\bar{S}_j^{cs,(2)}(b_T)\ &= \Bigl( \frac{\mu b_T}{b_0}\Bigr)^{\!4\epsilon} \Biggl[ \Bigl( \frac{\nu b_T R_j}{b_0}\Bigr)^{\!\eta} 
C_F C_A \bar{v}_A^{\rm in} +  R^{-4\epsilon} C_F C_A \bar{v}_A^{\rm out} \nn \\
&\quad+ \Bigl( \frac{\nu b_T R_j}{b_0}\Bigr)^{\!\eta} 
\Bigl[
\Bigl( \frac{\nu b_T R_j}{b_0}\Bigr)^{\!\eta} C_F^2 \bar{h}_{2F} + C_F C_A \bar{h}_{A} + C_F n_f T_F \bar{h}_{f}\Bigr] \notag \\
&\quad+  \Bigl( \frac{\nu b_T R_j}{b_0}\Bigr)^{\!\eta} R^{-2\epsilon}
C_F^2 \bar{p}_{2F} +  
R^{-4\epsilon}( C_F C_A \bar{p}_A + C_F n_f T_F \bar{p}_f) + R^{-2\epsilon} C_F C_A \bar{p}_\text{NGL} \notag \\
&\quad+   R^{-4\epsilon} \bigl(C_F^2 \bar{g}_{2F} + C_F C_A \bar{g}_A + C_F n_f T_F \bar{g}_f\bigr)\Biggr],
\end{align}
The first line of Eq.~\eqref{eq:bare_coft_phibavg} represents the bare real-virtual correction, while the remaining three lines correspond to double real emissions: The second line describes the case where both emissions are inside the jet cone labeled by $j$. The third and fourth lines correspond to configurations with one radiation inside and one outside the jet cone, and with both emissions outside the jet cone, respectively. 
Note in particular that the dependence on $R$ and $\eta_j$ is fully analytic, the various auxiliary functions introduced here only contain numbers and regulator powers.

The real-virtual contributions are of NLO form:
\begin{align}
    \bar{v}_A^{\rm in}
    &=  - \frac{1}{\epsilon^4} + \frac{4}{\eta \epsilon^3}  - \frac{16 \ln{2}}{\eta\epsilon^2} + \frac{\frac{7 \pi^2}{6}  + 8 \ln^2 2}{\epsilon^2}  + \frac{2 \left(\pi^2 + 16 \ln^2 2\right)}{\eta\epsilon}  - \frac{2\left(10 \pi^2 \ln{2} + 32 \ln^3{2} + 34 \zeta_3\right)}{3\epsilon}\notag\\
    &\quad  -\frac{4 \left(6 \pi^2 \ln{2} + 32 \ln^3{2} + 46 \zeta_3\right)}{3 \eta}     + \frac{1009 \pi^4}{360} + \frac{52}{3} \pi^2 \ln^2{2} + 32 \ln^4{2} + 152 \zeta_3\ln{2}\\
    \bar{v}_A^{\rm out} 
    &= \frac{1}{\epsilon^4} - \frac{\pi^2}{2\epsilon^2} + \frac{8\zeta_3}{3\epsilon} +
     \frac{\pi^4}{120}
\end{align}
The $C_F^2$ terms arise from uncorrelated double emissions, which can be derived directly from the NLO collinear-soft function via Non-Abelian Exponentiation (NAE), followed by averaging over $\phi_b$,
\begin{align}
    \bar{h}_{2F} 
    &= \frac{6}{\epsilon^4} - \frac{8 \ln{2}}{\epsilon^3} - \frac{\pi^2}{3 \epsilon^2}  + \frac{4 \pi^2 \ln{2}}{3 \epsilon}  + \frac{1}{\eta^2} \left(\frac{8}{\epsilon^2} - \frac{32 \ln{2}}{\epsilon} + 64 \ln^2{2} + \frac{20 \pi^2}{3}\right)  \notag\\
    &\quad  +  \frac{1}{\eta} \left[-\frac{8}{\epsilon^3} + \frac{16 \ln{2}}{\epsilon^2} - \frac{4}{3} \left(10 \pi^2 \ln{2} + 32 \ln^3{2} + 46 \zeta_3\right)\right]  
    +\frac{301 \pi^4}{180} + \frac{32 \pi^2 \ln^2{2}}{3} +\frac{4}{3} \left(16 \ln^4{2} + 92 \zeta_3\ln{2} \right), \\
    \bar{p}_{2F} 
    &= - \frac{4}{\epsilon^4}  + \frac{4 \left(\pi^2 + 6 \ln^2{2}\right)}{3 \epsilon^2}  - \frac{8 \left(\pi^2 \ln{2} + 4 \ln^3{2} + 4\zeta_3\right)}{3 \epsilon}\notag \\
    &\quad + \frac{1}{\eta} \left[ \frac{8}{\epsilon^3} - \frac{16 \ln{2}}{\epsilon^2} + \frac{16 \ln^2{2}}{\epsilon} + \frac{16}{3} \left(-2 \ln^3{2} + \zeta_3\right) \right] 
    +\frac{16 \pi^4}{45} + \frac{8}{3} \pi^2 \ln^2{2} + 8 \left(\ln^4{2} + 2 \zeta_3 \ln{2} \right),\\
    \bar{g}_{2F} 
    &= \frac{2}{\epsilon^4} - \frac{5\pi^2}{3\epsilon^2} - \frac{68\zeta_3}{3\epsilon} - \frac{101 \pi^4}{180}.
\end{align}
We adopt the parametrization of Ref.~\cite{Bell:2018oqa} to compute the correlated double emissions pole by pole, and derive the following results for double in-cone radiation:
\begin{align}
    \bar{h}_{A} 
    &= \frac{1}{\epsilon^4} + \frac{11}{6\epsilon^3} + \left( \frac{67}{18} - \frac{5 \pi^2}{3} - 8 \ln^2 2 \right) \frac{1}{\epsilon^2} + \left( \frac{211}{27} - \frac{121 \pi^2}{36} + 8 \pi^2 \ln2 - \frac{44 \ln^2 2}{3} + \frac{64 \ln^3 2}{3} + \frac{35 \zeta_3}{3} \right) \frac{1}{\epsilon} \notag\\
&\quad + \biggl[- \frac{4}{\epsilon^3} + \left(-\frac{22}{3} + 16 \ln2\right)\frac{1}{\epsilon^2} + \left( -\frac{134}{9} - \frac{4 \pi^2}{3} + \frac{88 \ln2}{3} - 32 \ln^2 2 \right) \frac{1}{\epsilon} \notag\\
&\quad + \left( -\frac{808}{27} - \frac{55 \pi^2}{9}  + \frac{536 \ln2}{9} + \frac{16 \pi^2 \ln2}{3} - \frac{176 \ln^2 2}{3} + \frac{128 \ln^3 2}{3}  + \frac{268 \zeta_3}{3} \right)\biggr] \frac{1}{\eta} -365.293(48), \\
    \bar{h}_{f} 
    &= -\frac{2}{3 \epsilon^3} - \frac{10}{9 \epsilon^2} + \left( -\frac{74}{27} + \frac{11 \pi^2}{9} + \frac{16 \ln^2 2}{3} \right) \frac{1}{\epsilon} + \biggl[\frac{8}{3 \epsilon^2} + \left(\frac{40}{9} - \frac{32 \ln2}{3}\right) \frac{1}{\epsilon}   \notag\\
&\quad \left(\frac{224}{27} + \frac{20 \pi^2}{9} - \frac{160 \ln2}{9} + \frac{64 \ln^2 2}{3}\right)\biggr]\frac{1}{\eta} -40.8677(14) 
\end{align}
The regulator poles in the equations above can be derived using consistency relations originating from anomalous dimensions, which we have used to replace our (compatible) numerical results with the corresponding analytic results. The finite matching correction is not subject to the consistency relations and thus remains numerical.

We can obtain the double out-cone results by transforming a same-hemisphere contribution of the NNLO hemisphere soft function in \cite{Kelley:2011ng} from momentum space to Fourier $\vec{b}_T$ space,
\begin{align}
    \bar{g}_{A} 
    &=-\frac{1}{\epsilon^4} - \frac{11}{6 \epsilon^3} + \frac{-67 + 6 \pi^2}{18 \epsilon^2} + \frac{-47840 - 2010 \pi^2 + 1233 \pi^4 - 75240 \zeta_3}{3240} + \frac{-772 - 33 \pi^2 + 468 \zeta_3}{108 \epsilon} \\
    \bar{g}_{f} 
    &= \frac{2}{3 \epsilon^3} + \frac{10}{9 \epsilon^2} + \frac{38 + 3 \pi^2}{27 \epsilon} + \frac{238 + 15 \pi^2 + 684 \zeta_3}{81}.
\end{align}

The contribution from correlated one-in/one-out emissions can be computed using either a method-of-regions approach or a dedicated EFT treatment involving a coft mode, and yields
 \begin{align}
    \bar{p}_f &= \frac{4(3-2\pi^2)}{9\epsilon} -\frac{68}{9} + \frac{64\pi^2}{27} -\frac{16\zeta_3}{3},  \\
    \bar{p}_A &= \left(-\frac{2}{3}+\frac{22\pi^2}{9} - 4\zeta_3\right)\frac{1}{\epsilon} + \frac{40}{9} -\frac{134\pi^2}{27} + \frac{44\zeta_3}{3} +\frac{8\pi^4}{45},\\
    \bar{p}_\text{NGL} &=\frac{2 \pi^2}{3 \epsilon^2} + \left(8 \zeta_3-\frac{4}{3} \pi^2 \ln 2\right)\frac{1}{\epsilon} + \left( \frac{13 \pi^4}{45} + \frac{4}{3} \pi^2 \ln^2 2 - 16 \zeta_3\ln 2 \right)
\end{align}
The function $\bar{p}_\text{NGL}$ here is an effective object that is useful solely for a slicing application.
When we analyze our setup in a Method-of-Regions setting, $\bar{p}_\text{NGL}$ encodes the contribution from a soft emission inside the jet, together with an emission outside the jet whose energy scale is suppressed by a power of $R$ compared to the in-jet emission. In an EFT treatment~\cite{Becher:2015hka,Becher:2016mmh,Becher:2016omr} these emissions would be described by two separate field modes: A collinear-soft mode with standard soft virtuality, and a coft mode with similar rapidity but additional virtuality suppression\footnote{There would then also be a third mode, a global soft mode with the same virtuality as the collinear-soft mode, but isotropic in terms of rapidity, which neither the collinear-soft nor coft modes are. This mode then gives rise to the global soft function.}. In EFT-based factorization different field modes are modeled by different ingredient functions, our function $\bar{p}_\text{NGL}$ is thus a simplified expression: It consists of the product of NLO collinear-soft and coft functions in the non-global factorization, fully integrated over the cross-talk between these functions. This is sufficient for fixed-order purposes, but would not be adequate if we wanted to resum the corresponding logarithms.

\end{document}